\documentclass[conference]{IEEEtran}
\IEEEoverridecommandlockouts
\usepackage{amsfonts}
\usepackage{multirow}
\usepackage{makecell}
\usepackage{mathrsfs}
\usepackage[ruled]{algorithm2e}
\usepackage{color} 
\usepackage{cite}
\usepackage{amsmath,amssymb,amsfonts}
\usepackage{algorithmic}
\usepackage{graphicx}
\usepackage{textcomp}
\usepackage{xcolor}
\usepackage{url}
\def\BibTeX{{\rm B\kern-.05em{\sc i\kern-.025em b}\kern-.08em
    T\kern-.1667em\lower.7ex\hbox{E}\kern-.125emX}}
\begin{document}

\title{Efficient Bearing Sensor Data Compression via an Asymmetrical Autoencoder with a Lifting Wavelet Transform Layer\\
\thanks{This work was supported by the National Science Foundation (NSF) under grant 2303700. AE Cetin is also funded by the DoE Award Number: DE-SC0023715.}
}

\author{\IEEEauthorblockN{ Xin Zhu}
\IEEEauthorblockA{\textit{Department of Electrical and Computer Engineering} \\
\textit{University of Illinois Chicago}\\
Chicago, IL, USA\\
xzhu61@uic.edu}
\and
\IEEEauthorblockN{ Ahmet Enis \c Cetin}
\IEEEauthorblockA{\textit{Department of Electrical and Computer Engineering} \\
\textit{University of Illinois Chicago}\\
Chicago, IL, USA \\
aecyy@uic.edu}

}

\maketitle

\begin{abstract}
Bearing data compression is vital to manage the large volumes of data generated during condition monitoring. In this paper, a novel asymmetrical autoencoder with a lifting wavelet transform (LWT) layer is developed to compress bearing sensor data. The encoder part of the network consists of a convolutional layer followed by a wavelet filterbank layer. 
Specifically, a dual-channel convolutional block with diverse convolutional kernel sizes and varying processing depths is integrated into the wavelet filterbank layer to enable comprehensive feature extraction from the wavelet domain.
Additionally, the adaptive hard-thresholding nonlinearity is applied to remove redundant components while denoising the primary wavelet coefficients. On the decoder side, inverse LWT, along with multiple linear layers and activation functions, is employed to reconstruct the original signals.
Furthermore, to enhance compression efficiency, a sparsity constraint is introduced during training to impose sparsity on the latent representations.
The experimental results demonstrate that the proposed approach achieves superior data compression performance compared to state-of-the-art methods.
\end{abstract}


\section{Introduction}
Bearings play a critical role in various mechanical systems, facilitating both rotational and linear motion with diminished friction~\cite{khaire2020role}. 
Monitoring and analyzing the condition of bearings is crucial for maintaining the reliability and operational efficiency of mechanical systems, as undetected bearing failures can result in substantial downtime, costly repairs, and potentially catastrophic system failures~\cite{peeters2018vibration}. However, condition monitoring techniques, such as vibration analysis and acoustic emission methods~\cite{ali2014acoustic,ali2016acoustic}, generate large volumes of data over time, which presents challenges for rapid transmission and real-time analysis in bandwidth-constrained environments.

Bearing data compression mitigates these challenges by minimizing the data volume while retaining essential information required for precise analysis and fault diagnosis. Vibration data compression methods can generally be divided into lossless and lossy compression techniques. 

Lossless compression methods reduce data size without any loss of information~\cite{hossain2023empirical,shi2023lossy,altamimi2024lossless}. Huang \textit{et al.} proposed a divide-and-compress lossless compression scheme~\cite{huang2015divide}.
The discrete cosine transform (DCT) is used to partition the bearing data into two segments: the first consists of a few DCT coefficients that capture the majority of the signal's energy, while the second part contains high-frequency DCT coefficients that represent widely dispersed differences with minimal energy. Next, distinct encoding strategies are implemented for each short-time segment according to their unique data properties. However, it has a low compression efficiency.

Lossy data compression techniques achieve higher compression ratios by eliminating redundancy, potentially resulting in a reduction in quality~\cite{yu2024vlsi,dao2015lossy,higgins2010eeg}. In many practical applications, there is no need to compress the data in a lossless manner \cite{luo2024aln,hamdan2024effect}.
In~\cite{yin2023three}, a three-dimensional vibration data compression method was developed by incorporating the dimensions of length, width, and height to enhance compression efficiency. 
Guo \textit{et al.} proposed an optimal ensemble empirical mode decomposition (EEMD) based bearing data compression approach. It extracts the intrinsic mode function associated with bearing faults and compresses this component, rather than the original signal, to improve compression ratio~\cite{guo2013novel}. Additionally, an asymmetrical autoencoder was designed to compress vibration signals\cite{zhu2024novel}. It introduces a sparsifying discrete cosine Stockwell transform layer to eliminate redundant data and capture important features in the latent space. The DCT-based lossy compression methods have a low reconstruction accuracy compared to the proposed wavelet transform-based method (this article). 

To address the challenges of low reconstruction accuracy in lossy compression methods and the inefficiency of lossless compression techniques, this paper presents an asymmetrical autoencoder with a lifting wavelet transform layer (AAELWTL) for compressing bearing sensor data. 
The encoder module is designed with low complexity to facilitate implementation on sensors, while the decoder module incorporates a more complex structure to improve reconstruction accuracy.
Specifically, a novel lifting wavelet transform (LWT) layer is introduced in the encoder to extract an efficient latent space representation.
The key idea of the LWT layer is to leverage convolutional layers to capture important frequency domain features while an adaptive hard-thresholding layer is applied to eliminate redundant data and appropriately scale (denoise) the large coefficients.
In the decoder, the inverse LWT (ILWT) is combined with three linear layers to reconstruct the original signals. 
Additionally, a sparsity penalty is integrated into the training process to enforce sparsity on the latent space coefficients, further optimizing the compression efficiency.
The proposed model exhibits superior compression efficiency and reconstruction accuracy compared to other state-of-the-art models, as demonstrated on two bearing datasets: the MFPT~\cite{mfpt2019} and XJTU~\cite{xjtu2019} datasets. 

\begin{figure}[tb]
\centerline{\includegraphics[scale=.45]{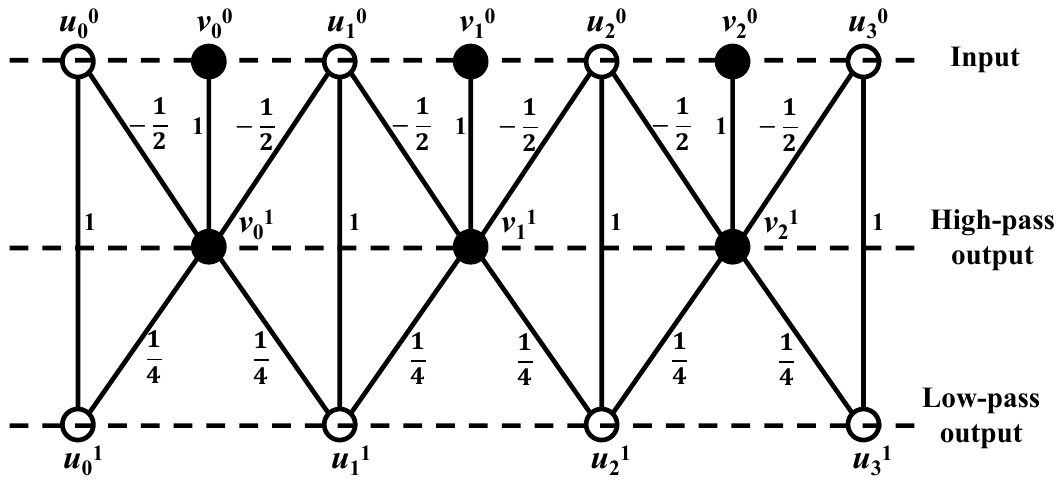}}
\caption{Lifting low-pass and high-pass filtering steps for the (5, 3) Daubechies biorthogonal filterbank.}
\label{fig: Lwavelet}
\end{figure}
\section{Preliminaries}
\label{sec:Background}

LWT is used in a wide range of applications, including the JPEG 2000 image compression standard \cite{kim1992class,
rabbani2002overview,
sweldens1996wavelets,
daubechies2005factoring,
gerek2000adaptive, gerek2005lossless, gerek20052}. In LWT, the low-pass and high-pass filtering operations are applied to the input after the lazy filterbank, which decomposes the input samples into even-indexed components, denoted as $u_i^{0}$, and odd-indexed components, denoted as $v_i^0$.
 For the (5, 3) filter bank~\cite{rabbani2002overview}, the highpass-filtered sub-signal $v_i^1$ is obtained by estimating each odd-indexed sample as a linear combination of two neighboring even-indexed samples and calculating the difference as follows:
\begin{equation}
    v_i^1= v_i^{0} - 0.5 ( u_i^0 +u_{i+1}^0)
\label{eq: wavelet1}
\end{equation}
where $i$ is the sample index. The low-pass filtered sub-signal $u_i^1$ is obtained using the update step:
\begin{equation}
    u_i^1 = u_i^0 + 0.25 ( v_i^1 +v_{i-1}^1)
\label{eq: wavelet2}
\end{equation}
This process is illustrated in Fig. \ref {fig: Lwavelet}.

\section{Asymmetrical autoencoder with a lifting wavelet transform layer}
The asymmetrical autoencoder is a variant of the traditional autoencoder architecture, where the encoder and decoder have different levels of complexity. In the realm of data compression, this asymmetry typically involves a more complex encoder designed to extract features and compress signals efficiently~\cite{kim2020efficient}. However, the complex encoder requires substantial computational resources. It poses a challenge for deploying autoencoders in low-cost sensor applications, where minimizing resource consumption is critical. 

\subsection{Structure of AAELWTL}
Unlike the established asymmetrical autoencoder~\cite{kim2020efficient}, the AAELWTL encoder is developed with low computational complexity for implementation in the sensor, while the decoder incorporates more sophisticated architectures to enhance data reconstruction accuracy. This approach is viable as the decoder operates on a high-performance host computer, capable of handling higher computational workloads than the sensor.
\begin{figure}[htbp]
\centerline{\includegraphics[scale=.55]{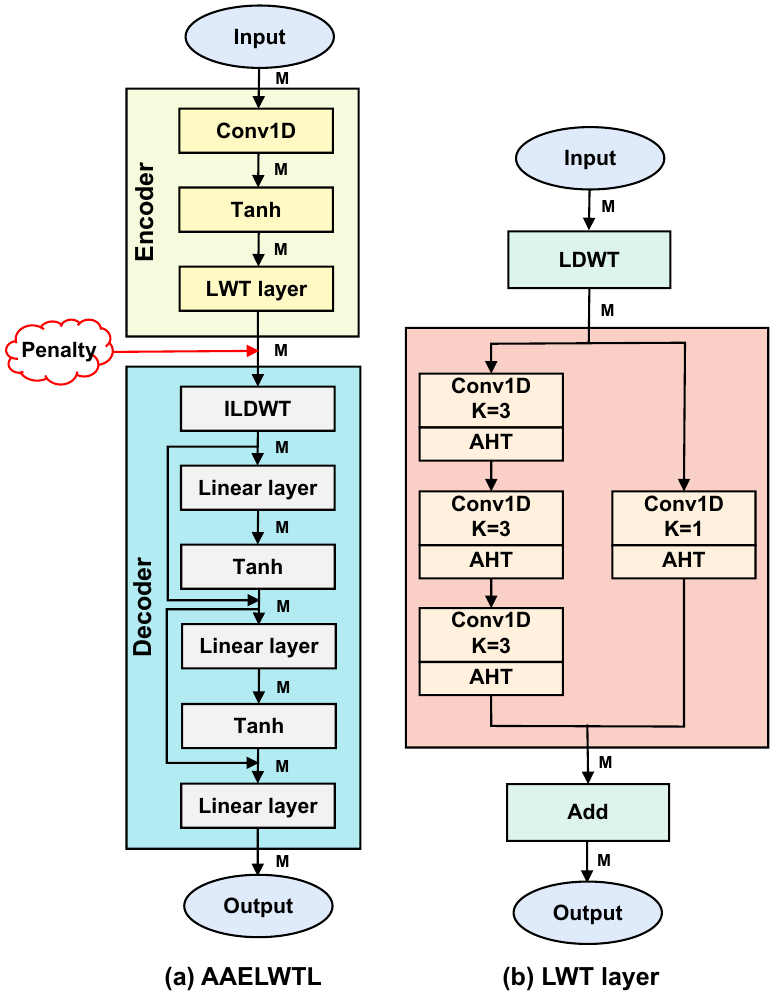}}
\caption{Block diagrams of the AAELWTL and LWT layer.}
\label{fig: AAELWTL}
\end{figure}

\noindent{\em Encoder Module of the AAELWTL}:

The bearing signal is divided into short-time segments, each with a duration of $M$ samples. Here, we choose $M=7$ because a small segment length minimizes computational cost and reduces the number of trainable parameters.
As is shown in Fig.~\ref{fig: AAELWTL}, each data segment is passed through a convolutional layer, followed by a tanh activation function. The output from the tanh function is denoted as $\mathbf{x}\in\mathbb{R}^M$. Next, the LWT of $\mathbf{x}$ is calculated as $\mathbf{X}\in\mathbb{R}^M$ using Eq.~(\ref{eq: wavelet1}) and Eq.~(\ref{eq: wavelet2}). In the wavelet domain, a convolutional block with two parallel branches is designed to extract features at different scales. The left branch consists of three consecutive 1D convolutional layers, each using a kernel size of 3. In contrast, the right branch features a one-by-one convolutional layer. 
The combination of different kernel sizes and processing depths allows for more comprehensive feature extraction from the wavelet domain.   

\begin{figure}[htbp]
\centerline{\includegraphics[scale=.25]{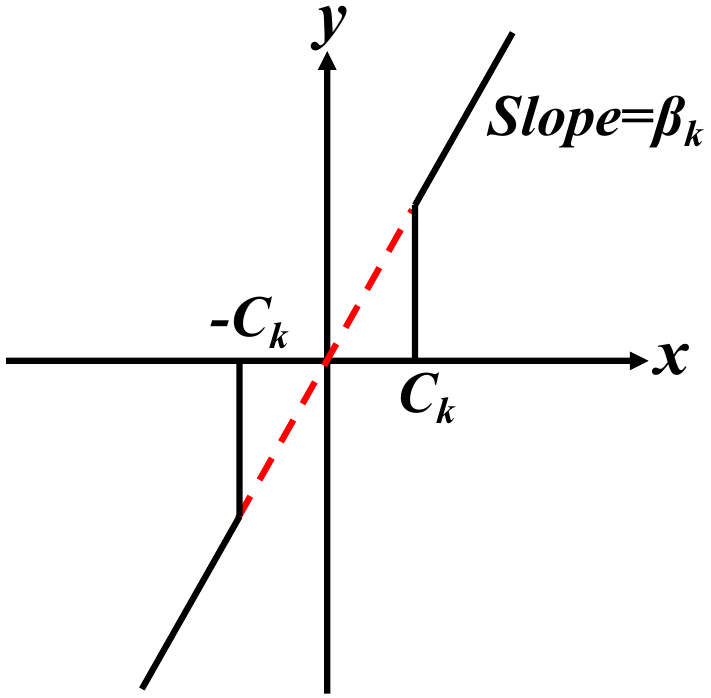}}
\caption{Adaptive hard-thresholding nonlinearity.}
\label{fig:hard}
\end{figure}

In the convolutional block, each convolutional layer is followed by an adaptive hard-thresholding (AHT) nonlinearity. Unlike the fixed threshold and slope used in conventional hard-thresholding (HT)~\cite{li2024ultrasound} and soft-thresholding (ST) functions~\cite{wang2024fourier}, the AHT layer introduces the trainable threshold $C_{k}$ and slope $\beta_{k}$ to adaptively suppress small entries and  
 scale (denoise) large coefficients. It is defined as:
\begin{equation}
    \begin{split}
        \widehat{X}_{k}= (\mathcal{E_{T}}(\widetilde{X}_k)+C_{k}\cdot\text{sign}(\mathcal{E_{T}}(\widetilde{X}_k)))\cdot \beta_k,
    \end{split}
    \label{Eq:HT}
\end{equation}
where $\mathcal{E_{T}}(\widetilde{X}_k)= \text{sign}(\widetilde{X}_k)\cdot\text{ReLU}(|\widetilde{X}_k|-C_{k})$ 
is the soft-thresholding function. 
$\widetilde{X}_k$ is the convolutional layer output.
$C_{k}$ and $\beta_{k}$ are optimized using the back-propagation algorithm~\cite{wang2023optimal,cilimkovic2015neural,lecun1988theoretical,van1992improving}, where $0\leq k \leq M-1$. The AHT is depicted in Fig.~\ref{fig:hard}.

\noindent{\em Decoder Part of the AAELWTL}:

The output of the LWT layer is quantized for transmission as described in Subsection~\ref{sec: transmission stage}. After receiving the quantized data, the decoder starts with the ILWT to convert data from the wavelet domain back to the time domain. The transformed data $\bar{\mathbf{x}}$ then passes through three separate linear layers and two tanh activation functions for bearing data reconstruction. Furthermore, a residual jump allows $\bar{\mathbf{x}}$ to bypass the intermediate layers and connect directly to the next linear layer, preserving crucial information and promoting more efficient training by mitigating vanishing gradient issues.

\subsection{ Offline model training}
\label{sec: training stage}
To improve compression efficiency, a sparsity penalty-based cost function is designed to enforce sparsity in the latent space coefficients. Assume the LWT layer output is $\mathbf{z}\in \mathbb{R}^M$. First, we define the activity of ${z}_{k}$ as follows:
\begin{equation}
    \hat{z}_{k}={\delta}(\mathbf{z})_{k}=\frac{e^{\lvert z_{k}\rvert}}{\sum_{i=0}^{M-1}e^{\lvert z_{i}\rvert}}, k=0,1,\cdots,M-1,
\end{equation}
where $\delta(\cdot)$ represents the softmax function. 
Given that the Kullback–Leibler divergence (KLD)~\cite{ng2011sparse,hershey2007approximating,perez2008kullback,belov2011distributions} quantifies the difference between two probability distributions, we utilize it as the sparsity regularization term:
\begin{equation}
\sum_{k=0}^{M-1} {\rm{KLD}}(\lambda\vert\vert\hat{z}_{k})
=\sum_{k=0}^{M-1}\lambda\log\frac{\lambda}{\hat{z}_{k}}+
    (1-\lambda)\log\frac{1-\lambda}{1-\hat{z}_{k}},
    \label{Eq:KL}
\end{equation}
where $\lambda$ stands for a sparsity parameter. Consequently, the total loss function is formulated as a weighted sum of the mean squared error and the KLD:
\begin{align}
&Loss=\frac{1}{M}\sum_{k=0}^{M-1} \left({x}_k-y_k\right)^2+\omega\sum_{k=0}^{M-1} {\rm{KLD}}\left(\lambda\vert\vert{\delta}(\mathbf{z})_k\right),
    \label{Eq:loss_p}
\end{align}
where $x_k$ and  $y_k$ denote the input and output element, respectively. $\omega$ represents the weight of the KLD term. Additionally, a threshold $\phi$ is set. During training, when the proportion of non-zero coefficients in $\mathbf{z}$ exceeds $\phi$, the training is terminated. This strategy helps eliminate small redundant coefficients.

\subsection{Binary Data Storage and/or Transmission}
\label{sec: transmission stage}
To optimize the transmission rate during the testing or transmission phase, the outputs of the LWT layer are converted from floating-point values to integers, as integers utilize fewer bits in memory than floating-point representations. The data conversion is computed as follows:
\begin{eqnarray}\label{Eq:code}
\tilde{\mathbf{z}}=\text{Round}(10^{\mu}\cdot {\mathbf{z}}/\alpha) ,   
\end{eqnarray}
where the parameters $\mu$ and $\alpha$ are utilized to control the compression ratio. $\text{Round}(\cdot)$ stands for the integer rounding function. Next, Run-Length Encoding (RLE)~\cite{akhter2010ecg} and Huffman Coding~\cite{sharma2010compression} are combined to enhance data compression and convert data into bitstreams.  RLE first simplifies the data by encoding consecutive zeroes as a single zero followed by the count of repetitions. Huffman Coding is subsequently applied to the RLE output, assigning variable-length codes based on the statistical frequency of symbols. After that, the encoded data is transmitted to the decoder for reconstruction.

\section{Experimental Results}
\label{Experimental Results}
In this work, the MFPT~\cite{mfpt2019} and XJTU~\cite{xjtu2019} datasets are employed to evaluate the bearing data compression performance of the proposed method. 
In the MFPT dataset, the record ``baseline\_1'' was selected for the compression experiment, with a sampling rate of 97,656 Hz. For the XJTU dataset, we used record ``bearing\_1\_1'', where the data was sampled at 25.6 kHz. 
To achieve a tradeoff between compression efficiency and reconstruction accuracy, $\mu$, $\alpha$, $\omega$ and $\phi$ as 3, 4, 10 and 0.6. Furthermore, the model is trained using a batch size of 30 with a learning rate of 0.001. Models are evaluated on the following metrics: the compression ratio (CR), percent root mean square difference (PRD), normalized percent root mean square difference (PRDN), root mean square error (RMSE), and quality score (QS)~\cite{jha2018electrocardiogram}. As the CR and QS rise, the compression efficiency and quality improve, while the disparity between the reconstructed signal and the original signal diminishes with a lower PRD, PRDN, and RMSE.

The sparse autoencoder (SAE)~\cite{ng2011sparse}, DCT perceptron (DCTP)~\cite{pan2024multichannel}, and asymmetrical autoencoder with a sparsifying discrete cosine Stockwell transform layer (AEDCST)~\cite{zhu2024novel} are selected as comparison methods given that they all have low complexity encoders for bearing data compression.

\begin{table}[ht]
\centering
\caption{Bearing data compression results. All methods are trained on the MFPT dataset. }
\begin{tabular*}{\linewidth}{@{\extracolsep{\fill}}|c|c|c|c|c|c|@{}}
\hline
\makecell{\bf{Test Data}}&\makecell{\bf{Metrics}}&\makecell{\bf{SAE }} &\makecell{\bf{DCTP}}& \makecell{\bf{AEDCST}}& \makecell{\bf{AAELWTL}} \\
\hline   
 \multirow{5}{*}{MFPT}
 &CR   &\makecell{7.57} & \makecell{9.23} & \makecell{9.73}& \makecell{\textbf{9.91}} \\
 &PRD   &\makecell{30.44} & \makecell{19.89} & \makecell{34.03}&\makecell{\textbf{17.29}} \\
 &PRDN   &\makecell{30.81} &\makecell{20.13} & \makecell{34.44}& \makecell{\textbf{17.50}} \\
 &RMSE   &\makecell{7.31} & \makecell{4.77} & \makecell{8.17}& \makecell{\textbf{4.15}} \\
 &QS   &{\makecell{0.25}} & \makecell{0.46} & \makecell{0.29}& \makecell{\textbf{0.57}} \\
\hline   
\multirow{5}{*}{XJTU}
 &CR   &\makecell{9.46} & \makecell{27.94} & \makecell{29.84}& \makecell{\textbf{31.37}} \\
 &PRD   &\makecell{29.89} & \makecell{22.72} & \makecell{32.93}& \makecell{\textbf{16.36}} \\
 &PRDN   &\makecell{29.89} & \makecell{22.72} & \makecell{32.93}& \makecell{\textbf{16.36}} \\
 &RMSE   &\makecell{7.64} & \makecell{5.81} &\makecell{8.41}& \makecell{\textbf{4.18}}\\
 &QS   &{\makecell{0.32}} & \makecell{1.23} &\makecell{0.91}& \makecell{\textbf{1.92}} \\
\hline
\end{tabular*}
\label{tab: experimental results}
\end{table}

\begin{figure}[b]
\centerline{\includegraphics[scale=.4]{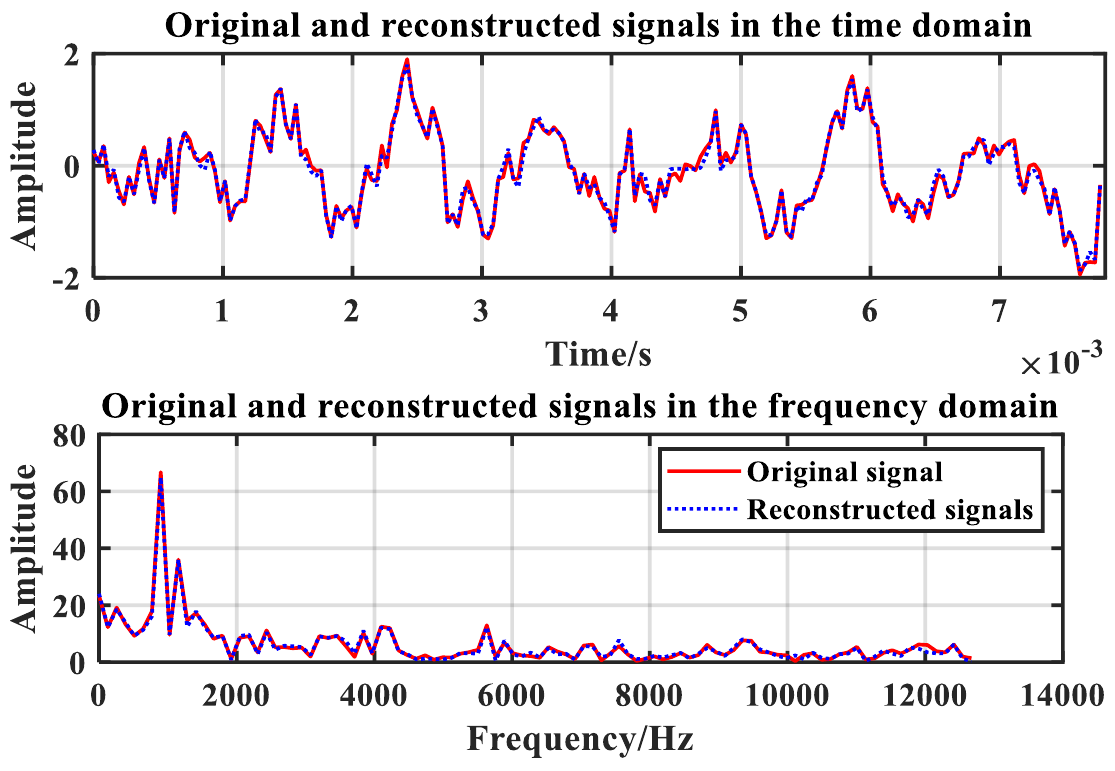}}
\caption{Comparison between the original data (red solid line) and the reconstructed data (blue dashed
line) in the time domain and the frequency domain on the XJTU dataset.}
\label{fig: compare}
\end{figure}

Table~\ref{tab: experimental results} summarizes the bearing data compression performance of the proposed approach versus the other three state-of-the-art methods on the MFPT and XJTU datasets. 
In this experiment, the top 20\% of the MFPT dataset is utilized for model training, while the models are evaluated on the XJTU dataset and the remaining 80\% of the MFPT dataset.
Compared to AEDCST, AAELWTL reduces PRD from 34.03 to 17.29 (49.19\%) and RMSE from 8.17 to 4.15 (49.20\%). The reason is that AAELWTL incorporates more linear layers and nonlinear activation functions in the decoder than AEDCST, resulting in improved data reconstruction accuracy. Additionally, AAELWTL is superior to DCTP in terms of a higher CR and a lower PRD. It is because AAELWTL implements an adaptive hard-thresholding layer, which effectively removes redundant data while appropriately scaling large coefficients within the wavelet domain. Therefore AAELWTL has a better compression efficiency. Besides, AAELWTL outperforms SAE. SAE works as a low-pass filter. However, AAELWTL not only preserves key coefficients in the low-frequency band but also retains significant values in the high-frequency band, as shown in Fig.~\ref{fig: compare}, thereby enhancing its reconstruction accuracy. 

Fig.~\ref{fig: compare} illustrates the comparison between original and reconstructed signals in both the time and frequency domains. In the time domain, the reconstructed signal closely aligns with the original signal. Similarly, in the frequency domain, both signals show strong overlap. This consistent performance across both domains demonstrates that AAELWTL effectively maintains the important temporal and frequency features of the signal during the data compression process.

Table~\ref{tab: Ablation} presents ablation experimental results, where different configurations are evaluated for their impact on compression performance. When the sparsity penalty and linear layers are removed from the proposed model, the performance deteriorates, indicating these components play a crucial role in enhancing the compression efficiency and reconstruction capability of AAELWTL. Moreover, AAELWTL with the AHT layer demonstrates superior compression performance compared to AAELWTL with either HT or ST layer. Although the ST layer can scale large coefficients, its fixed scaling factor restricts its adaptability to varying data characteristics. In contrast, the AHT layer introduces a trainable slope, allowing for dynamic adjustment of the data. Additionally, the original HT layer lacks the capability to scale large entries, further limiting its compression efficiency. 
\begin{table}[tb]
\centering
\caption{Ablation experimental result on the MFPT dataset. }
\begin{tabular*}{\linewidth}{@{\extracolsep{\fill}}|c|c|c|c|c|c|@{}}
\hline
\makecell{\bf{Metrics}}&\makecell{\bf{No penalty}}&\makecell{\bf{No linear}} &\makecell{\bf{ST}}& \makecell{\bf{HT}}& \makecell{\bf{AAELWTL}} \\
\hline   
 CR   &\makecell{9.80} & \makecell{9.90} & \makecell{9.67}& \makecell{9.84}& \makecell{\textbf{9.91}} \\
 PRD   &\makecell{17.38} &\makecell{19.33} & \makecell{42.32}&\makecell{17.30}& \makecell{\textbf{17.29}} \\
 PRDN   &\makecell{17.60} &\makecell{19.57} & \makecell{42.84}& \makecell{17.51}& \makecell{\textbf{17.50}} \\
 RMSE   &\makecell{4.17} & \makecell{4.64} & \makecell{10.16}& \makecell{4.15}& \makecell{\textbf{4.15}} \\
 QS   &\makecell{0.56} & \makecell{0.51} & \makecell{0.23}& \makecell{0.57}& \makecell{\textbf{0.57}} \\
\hline
\end{tabular*}
\label{tab: Ablation}
\end{table}

Table~\ref{tab: MACs} compares the computational complexity (measured in Multiply–Accumulates (MACs)) and the number of trainable parameters on the encoder side. Among all the models, AAELWTL exhibits the lowest MACs and the fewest trainable parameters. The reduced computational complexity and parameter count contribute to lower memory consumption and hardware costs, making the model more feasible for deployment in resource-limited sensors.

\begin{table}[tb]
\centering
\caption{The MACs and number of trainable parameters.}\label{table: MACs}
\begin{tabular}{|p{2cm}|p{2cm}|p{2cm}|}
\hline
\makecell{\bf{Algorithm}}&\makecell{\bf{MACs}}&\makecell{\bf{Parameters}} \\
\hline   
 \makecell{SAE}   &\makecell{107,520}&\makecell{3,120} \\
 \makecell{DCTP}   &\makecell{147,840}&\makecell{130} \\
 \makecell{AEDCST}   &\makecell{160,496}&\makecell{5,312} \\
 \makecell{Proposed method}   &\makecell{\textbf{60,480}}&\makecell{\textbf{74}} \\
\hline
\end{tabular}
\label{tab: MACs}
\end{table}

\section{Conclusion}
This study proposes an asymmetrical autoencoder with an LWT layer for bearing sensor data compression. 
A single convolutional layer 
cannot be trained in conjunction with an adaptive hard-thresholding nonlinearity. 
By imposing the convolutional layer after the fixed LWT, we can simultaneously train both the thresholds and the convolutional layer, allowing the model to adapt to the data while enhancing the compression ability of the encoder module. Furthermore, the decoder features a more sophisticated architecture, consisting of the ILWT, three linear layers, and two tanh activation functions, which significantly improve its reconstruction performance. The experimental results demonstrate that the proposed method surpasses other state-of-the-art models in terms of CR, PRD, PRDN, RMSE, and QS. Moreover, the AAELWTL encoder has a low computational complexity and a lower number of trainable parameters, making it well-suited for deployment on sensors to efficiently compress bearing data.







\bibliographystyle{IEEEbib}
\bibliography{strings}

\end{document}